\documentclass[12pt]{article}
\usepackage{epsf}
\usepackage{amsmath,amssymb}

\usepackage[dvips]{graphicx}

\usepackage{comment}

\begin{document}

\newcommand{\lsim}{\stackrel{<}{_\sim}}
\newcommand{\gsim}{\stackrel{>}{_\sim}}

\newcommand{\rem}[1]{{$\spadesuit$\bf #1$\spadesuit$}}

\renewcommand{\thefootnote}{\fnsymbol{footnote}}
\setcounter{footnote}{0}

\begin{titlepage}

\def\thefootnote{\fnsymbol{footnote}}

\begin{center}

\hfill UT-13-08\\
\hfill March, 2013\\

\vskip .75in

{\Large \bf 

  Probing Supersymmetric Model with Heavy Sfermions\\
  Using Leptonic Flavor and CP Violations

}

\vskip .75in

{\large Takeo Moroi and Minoru Nagai }

\vskip 0.25in

{\em Department of Physics, University of Tokyo,
Tokyo 113-0033, Japan}

\end{center}
\vskip .5in

\begin{abstract}

  We study leptonic flavor and CP violating observables in
  supersymmetric (SUSY) models with heavy sfermions, which is
  motivated by the recent results of the LHC experiments (i.e., the
  discovery of the Higgs-like boson with the mass of about $126\ {\rm
    GeV}$ and the negative searches for the superparticles).  Even if
  the sfermion masses are of $O(10-100\ {\rm TeV})$, signals may be
  within the reach of future leptonic flavor- and CP-violation
  experiments assuming that the off-diagonal elements of the sfermion
  mass matrices are unsuppressed compared to the diagonal ones.  We
  also consider the SUSY contribution to the $K^0$-$\bar{K}^0$ mixing
  parameters; we show that the leptonic observables can become as powerful
  as those in $K^0$-$\bar{K}^0$ mixing to constrain SUSY models.

\end{abstract}

\end{titlepage}

\renewcommand{\thepage}{\arabic{page}}
\setcounter{page}{1}
\renewcommand{\thefootnote}{\#\arabic{footnote}}
\setcounter{footnote}{0}

Recent progresses of the searches for new particles at the LHC have
provided important information about the physics at the electroweak
scale and beyond.  In particular, supersymmetry (SUSY), which is one
of the important candidates of the physics beyond the standard model,
has been seriously constrained by the results of the LHC, i.e, the
discovery of the Higgs-like boson with the mass of about $126\ {\rm
  GeV}$ \cite{Aad:2012gk, Chatrchyan:2012gu}, and the negative
searches for superparticles \cite{Aad:2012fqa,Chatrchyan:2012jx}.  In
particular, it is notable that the Higgs mass is preferred to close to
the $Z$-boson mass in the minimal SUSY standard model (MSSM) and that
the lightest Higgs mass of about $126\ {\rm GeV}$ is hardly realized
in the MSSM unless stops are heavier than $10\ {\rm TeV}$ or the
tri-linear scalar coupling constant of stops is enhanced.

These facts suggest a class of supersymmetric models, i.e., models
with heavy sfermions.  If the SUSY is broken with the SUSY breaking
scale corresponding to $m_{3/2}\sim 10-100\ {\rm TeV}$ (with $m_{3/2}$
being the gravitino mass), all the scalars in the MSSM sector (except
for the lightest Higgs boson) naturally have a mass of $O(10-100\ {\rm
  TeV})$.  Such a model is phenomenologically viable; the lightest
Higgs mass can be pushed up to $126\ {\rm GeV}$, while all the
superparticles (in particular, sfermions) can be out of the reach of
the LHC experiments.  If the gravitino mass is as heavy as
$m_{3/2}\sim 10-100\ {\rm TeV}$, it also helps to avoid the serious
cosmological gravitino problem \cite{Kawasaki:2008qe}.  In addition,
SUSY model with sfermion masses of $O(10-100\ {\rm TeV})$ is
compatible with the grand unified theory (GUT).  Thus, the scenario
with heavy sfermions has been recently attracted attentions
\cite{Giudice:2011cg, Ibe:2011aa, Moroi:2011ab, Kane:2011kj,
  ArkaniHamed:2012gw, Hall:2013uga, Ibe:2006de, Arganda:2012qp, Arganda:2013ve}.

Heavy sfermions are also advantageous to avoid (or to relax)
constraints from flavor and CP violations.  In SUSY models in which
the masses of superparticles are around $1\ {\rm TeV}$, it is often
the case that too large flavor and CP violations are induced by loop
diagrams with superparticles inside the loop.  With heavy enough
sfermions, such constraints are supposed to be avoided.  However, even
if the sfermion masses are around $10-100\ {\rm TeV}$, it is
non-trivial whether the model evades all the flavor and CP
constraints.  It is well known that the constraint from the CP
violation in the kaon decay (i.e., the constraint from the so-called
$\epsilon_K$ parameter) often gives the most stringent constraint and
that the SUSY contribution to $\epsilon_K$ may become larger than the
standard-model prediction even with the sfermion masses of $O(10-100\
{\rm TeV})$ \cite{Gabbiani:1996hi}.

The purpose of this letter is to reconsider the flavor and CP
constraints on the SUSY model, paying particular attention to the
heavy sfermion scenario.  We will see that, in some class of
well-motivated model, the constraint from $\epsilon_K$ is relaxed and
that lepton-flavor-violation (LFV) and CP violation may be also
powerful tools to study heavy sfermion scenario.  Future experiments
measuring $Br(\mu\rightarrow e\gamma)$, $Br(\mu\rightarrow 3e)$,
$\mu$-$e$ conversion rate, and the electron electric dipole moment
(EDM) $d_e$ will cover the parameter region with the sfermion masses
of $O(10-100\ {\rm TeV})$.

Let us start our discussion with introducing the framework of the
model of our interest.  In this letter, we consider the case where
SUSY is dynamically broken by the condensation of a chiral superfield
$Z$.  (There may exist more than one chiral superfields responsible
for the SUSY breaking, but the following discussion is unaffected.)
Allowing higher dimensional operators suppressed by the Planck scale
$M_{\rm Pl}\simeq 2.4\times 10^{18}\ {\rm GeV}$, the K\"ahler
potential may contain the following term
\begin{align}
  K \ni \frac{\kappa_{\Phi,IJ}}{M_{\rm Pl}^2} Z^\dagger Z 
  \Phi_I^\dagger \Phi_J,
\end{align}
where $\Phi$ denotes chiral superfields in the MSSM sector,
corresponding to $q_L({\bf 3},{\bf 2},1/6)$, $u_R^c(\bar{\bf 3},{\bf
  1},-2/3)$, $d_R^c(\bar{\bf 3},{\bf 1},1/3)$, $l_L({\bf 1},{\bf
  2},-1/2)$, and $e_R^c({\bf 1},{\bf 1},1)$ (with the gauge quantum
numbers for $SU(3)_C$, $SU(2)_L$ and $U(1)_Y$ being shown in the
parenthesis), $\kappa_{\Phi,IJ}$ is a constant, and $I$ and $J$ are
generation indices which run from $1$ to $3$.  After the SUSY
breaking, the $F$-component of $Z$ acquires a vacuum expectation
value, and the soft SUSY breaking mass squareds of sfermions in the
MSSM sector show up.  In this letter, we assume that the SUSY is
broken with relatively large value of the gravitino mass of
$m_{3/2}\sim O(10-100\ {\rm TeV})$.  Assuming that
$\kappa_{\Phi,IJ}\sim O(0.1-1)$, sfermion masses (as well as the SUSY
breaking Higgs mass parameters) are of the same order.  In this
framework, the K\"ahler potential may contain the terms like
\begin{align}
  K \ni c_1 H_{u}H_{d} + 
  \frac{c_2}{M_{\rm Pl}^{2}} Z^{\dagger} Z H_{u}H_{d} + {\rm h.c.},
\end{align}
where $H_u$ and $H_d$ are up- and down-type Higgses, respectively.
With $c_1$ and $c_2$ being $O(0.1-1)$, we expect that the SUSY
invariant Higgs mass (so-called $\mu$ parameter) and the bi-linear
SUSY breaking parameter (so-called $B$ parameter) are both expected to
be of the same order of the gravitino mass \cite{Giudice:1988yz,
  Inoue:1991rk, Ibe:2006de}.

Even if the sfermion masses are of $O(10-100\ {\rm TeV})$, the gaugino
masses are model-dependent.  If the SUSY breaking sector contains a
singlet field, then the gaugino masses may arise from the direct
$F$-term interaction between the gauge multiplet and the SUSY breaking
field.  In such a case, we expect that the gaugino masses are of the
order of the gravitino mass.  We call such a case as heavy gaugino
case, in which we assume that the gaugino masses $M_A$ obey the simple
GUT relation:
\begin{eqnarray}
  \frac{3}{5}
  \frac{M_1}{g_1^2} = 
  \frac{M_2}{g_2^2} = 
  \frac{M_3}{g_3^2},
\end{eqnarray}
where $g_{A}$ denote standard model gauge coupling constants.  (Here,
$A=1$, $2$ and $3$ correspond to $U(1)_Y$, $SU(2)_L$ and $SU(3)_C$,
respectively.)  On the contrary, if there is no singlet field, effect
of the anomaly-mediated SUSY breaking (AMSB) may dominate the gaugino
masses.  In such a case, the gaugino masses are obtained as
\cite{Randall:1998uk, Giudice:1998xp}\footnote
{If the $\mu$ parameter is as large as the gravitino mass, the gaugino
  masses may not obey the simple anomaly-mediation relation
  \cite{Giudice:1998xp, Ibe:2006de}.  Here, we assume that such an
  effect is negligible.}
\begin{align}
  M_{A}^{\rm (AMSB)}
  = -\frac{b_A g_A^{2}}{16\pi^{2}} m_{3/2},
  \label{GauginoMass}
\end{align}
where $b_{A}$ denote coefficients of the renormalization-group (RG)
equations for $g_{A}$, i.e., $b_{A}= (-11, -1, 3)$.\footnote
{The AMSB scenario without singlet field is advantageous for cosmology
  because the lightest neutralino (which may be the neutral Wino) can
  be a good candidate of dark matter \cite{Giudice:1998xp,
    Moroi:1999zb}, and also because the cosmological problem due to
  the late-time decay of the SUSY breaking field (i.e., the so-called
  Polonyi problem) may be avoided \cite{Banks:1993en}.}
In the following discussion, we consider both cases.  We also comment
here on the tri-linear scalar couplings (i.e., so-called $A$
parameters); in the following discussion, their effects are not
important, so are neglected.

We parameterize the soft SUSY breaking terms of sfermions in the gauge
basis as
\begin{align}
  {\cal L}_{\rm soft} =& 
  \tilde{q}_{L,I}^\dagger {\cal M}^2_{\tilde{q}_{L},IJ}
  \tilde{q}_{L,J}
  + \tilde{u}_{R,I}^{c} {\cal M}^2_{\tilde{u}_{R},IJ}
  \tilde{u}_{R,J}^{c\dagger}
  + \tilde{d}_{R,I}^{c} {\cal M}^2_{\tilde{d}_{R},IJ}
  \tilde{d}_{R,J}^{c\dagger}
  \nonumber \\ &
  + \tilde{l}_{L,I}^\dagger {\cal M}^2_{\tilde{l}_{L},IJ}
  \tilde{l}_{L,J}
  + \tilde{e}_{R,I}^{c} {\cal M}^2_{\tilde{e}_{R},IJ}
  \tilde{e}_{R,J}^{c\dagger},
\end{align}
or in the flavor basis,
\begin{align}
  {\cal L}_{\rm soft} =& 
  \tilde{u}_{L,i}^\dagger {\cal M}^2_{\tilde{u}_{L},ij}
  \tilde{u}_{L,j}
  + \tilde{d}_{L,i}^\dagger {\cal M}^2_{\tilde{d}_{L},ij}
  \tilde{d}_{L,j}
  + \tilde{u}_{R,i}^{c} {\cal M}^2_{\tilde{u}_{R},ij}
  \tilde{u}_{R,j}^{c\dagger}
  + \tilde{d}_{R,i}^{c} {\cal M}^2_{\tilde{d}_{R},ij}
  \tilde{d}_{R,j}^{c\dagger}
  \nonumber \\ &
  + \tilde{\nu}_{L,i}^\dagger {\cal M}^2_{\tilde{\nu}_{L},ij}
  \tilde{\nu}_{L,j}
  + \tilde{e}_{L,i}^\dagger {\cal M}^2_{\tilde{e}_{L},ij}
  \tilde{e}_{L,j}
  + \tilde{e}_{R,i}^{c} {\cal M}^2_{\tilde{e}_{R},ij}
  \tilde{e}_{R,j}^{c\dagger}.
  \label{L_soft}
\end{align}
(Here and hereafter, the tilde is for superpartners of the
standard-model particles.)  In Eq.\ \eqref{L_soft}, all the mass
matrices are in the basis in which the fermion mass matrices are
diagonalized; thus, ${\cal M}^2_{\tilde{u}_{L}}$ and ${\cal
  M}^2_{\tilde{d}_{L}}$ are related by the CKM matrix.  It is obvious
that the non-vanishing values of ${\cal M}^2_{\tilde{f},ij}$ (with
$i\neq j$) become new sources of flavor violation.  In addition, the
off-diagonal elements of the mass matrices can have non-vanishing
phases, and are new sources of CP violation.

For the following discussion, it is convenient to define
\begin{align}
  \Delta_{\tilde{f},ij} \equiv 
  \frac{{\cal M}^2_{\tilde{f},ij}-m^2_{\tilde{f}}\delta_{ij}}
  {m^2_{\tilde{f}}},
\end{align}
where
\begin{align}
  m^2_{\tilde{f}} \equiv
  {\cal M}^2_{\tilde{f},11}.
\end{align}
The value of $\Delta_{\tilde{f},ij}$ is model-dependent.  If all the
higher-dimensional operators suppressed by the Planck scale are
allowed, $\Delta_{\tilde{f},ij}\sim O(0.1)$ is expected.\footnote
{One might think $\Delta_{\tilde{f},ij}$ is naturally of $\sim O(1)$.
  However, if the off-diagonal elements are larger than the diagonal
  ones, there may exist a negative eigenvalue of ${\cal
    M}^2_{\tilde{f},ij}$, which results in color and/or charge
  breaking vacuum.  Thus, in order to maintain the $SU(3)_C\times
  U(1)_{\rm em}$ symmetry, we assume $\Delta_{\tilde{f},ij}\sim
  O(0.1)$.}

Now, we discuss the rates of LFV.  As we have mentioned, non-vanishing
values of $\Delta_{\tilde{e}_L,ij}$, $\Delta_{\tilde{e}_R,ij}$, and
$\Delta_{\tilde{\nu}_L,ij}$ induce various LFV processes.  With
$\Delta_{\tilde{e}_L,ij}$, $\Delta_{\tilde{e}_R,ij}$, and
$\Delta_{\tilde{\nu}_L,ij}$ being fixed, the LFV rates become smaller
as the sleptons become heavier.  To see the mass scale of sleptons
accessible with the experiments, we simply assume the following
structure of the slepton mass matrix:
\begin{align}
  {\cal M}^2_{\tilde{e}_{R},ij} =&
  m_{\tilde{l}}^2 \left( \delta_{ij} + \Delta_{\tilde{e}_R,ij} \right),
  \label{m2er}
  \\
  {\cal M}^2_{\tilde{e}_{L},ij} =& 
  m_{\tilde{l}}^2 \left( \delta_{ij} + \Delta_{\tilde{l}_L,ij} \right),
  \\
  {\cal M}^2_{\tilde{\nu}_{L},ij} =& 
  m_{\tilde{l}}^2 \left( \delta_{ij} + \Delta_{\tilde{l}_L,ij} \right),
  \label{m2nl}
\end{align}
and calculate the LFV rate; the detailed formulae of the LFV rates can
be found in \cite{Hisano:1995cp}.

\begin{figure}[t]
  \centerline{\epsfxsize=0.8\textwidth\epsfbox{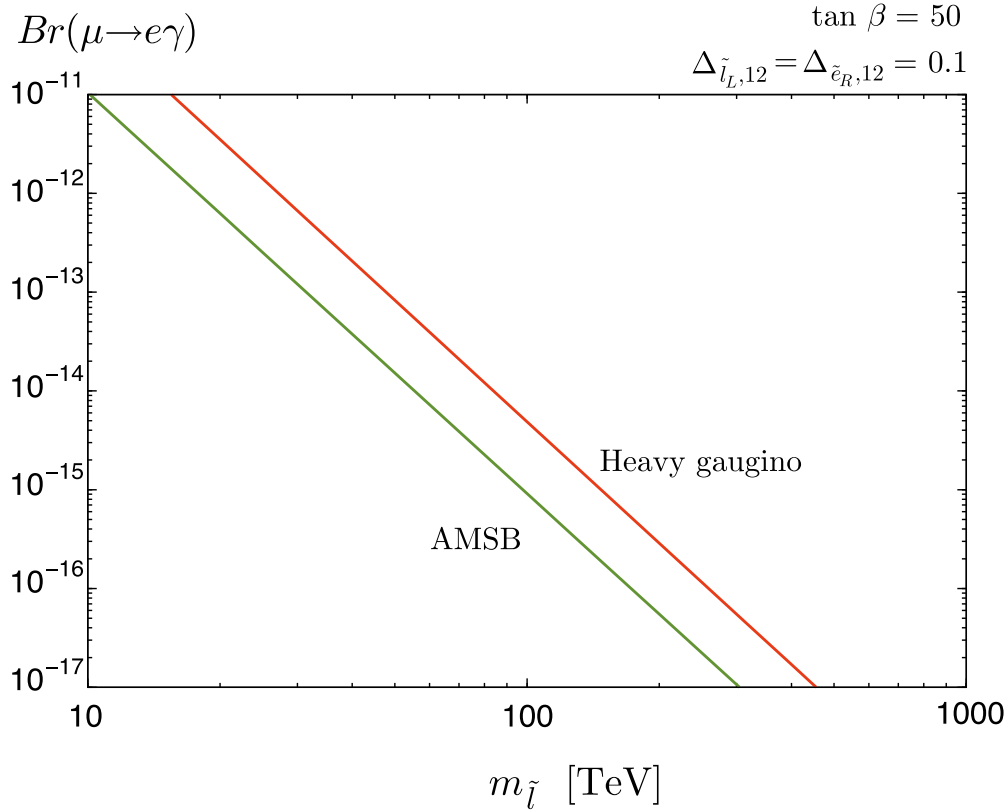}}
  \caption{$Br(\mu\rightarrow e\gamma$) as a function of the slepton
    mass $m_{\tilde{l}}$ for $\tan\beta=50$ and $\mu=m_{\tilde{l}}$.
    In addition, $\Delta_{\tilde{l}_L,12}=\Delta_{\tilde{l}_L,21}=
    \Delta_{\tilde{e}_R,12}=\Delta_{\tilde{e}_R,21}=0.1$, while other
    components of $\Delta_{\tilde{f},ij}$ are taken to be zero.  Upper
    (red) and lower (green) lines are for the heavy gaugino case with
    $M_3=m_{\tilde{l}}$ and the AMSB case with
    $m_{3/2}=5m_{\tilde{l}}$, respectively.}
  \label{fig:brMEG}
\end{figure}

We first consider the $\mu\rightarrow e\gamma$ process.  The simplest
possibility to induce the $\mu\rightarrow e\gamma$ process is to
introduce non-vanishing values of $\Delta_{\tilde{l}_L,12}$ and/or
$\Delta_{\tilde{e}_R,12}$.  In Fig.\ \ref{fig:brMEG}, we plot
$Br(\mu\rightarrow e\gamma)$ in such a case as a function of the
slepton mass $m_{\tilde{l}}$.  Here, we take
$\Delta_{\tilde{l}_L,12}=\Delta_{\tilde{l}_L,21}=
\Delta_{\tilde{e}_R,12}=\Delta_{\tilde{e}_R,21}=0.1$.  (Other
components of $\Delta_{\tilde{f},ij}$ are taken to be zero; notice
that, in such a case, $Br(\mu\rightarrow e\gamma)$ is approximately
proportional to $\Delta_{\tilde{l},12}^2$.)  In order to see how large
$Br(\mu\rightarrow e\gamma)$ can be, we adopt relatively large value
of $\tan\beta$ (which is the ratio of up- and down-type Higgs bosons);
in our numerical calculation, we take $\tan\beta=50$.  (For the case
of large $\tan\beta$, $Br(\mu\rightarrow e\gamma)$ is approximately
proportional to $\tan^2\beta$.)  For the gaugino mass, we consider two
cases: the heavy gaugino case with GUT relation (with
$M_3=m_{\tilde{l}}$) and the AMSB case (with
$m_{3/2}=5m_{\tilde{l}}$).\footnote
{For the AMSB case, if we naively take $m_{3/2}=m_{\tilde{l}}$, the
  gluino mass may conflict with the LHC bounds in some parameter
  region of our study below.  Thus, we assume a slight suppression of
  the slepton mass relative to the gravitino mass in the AMSB case.}
In addition, we take $\mu=m_{\tilde{l}}$.  We can see that
$Br(\mu\rightarrow e\gamma)$ becomes smaller in the AMSB case.  This
is because, in the case of large $\tan\beta$, the $\mu\rightarrow
e\gamma$ process is induced by diagrams with chirality-flip due to the
gaugino mass.  Thus, the amplitude for the AMSB case is suppressed by
the factor of about $M_{1,2}/m_{\tilde{l}}$.

\begin{figure}[t]
  \centerline{\epsfxsize=0.8\textwidth\epsfbox{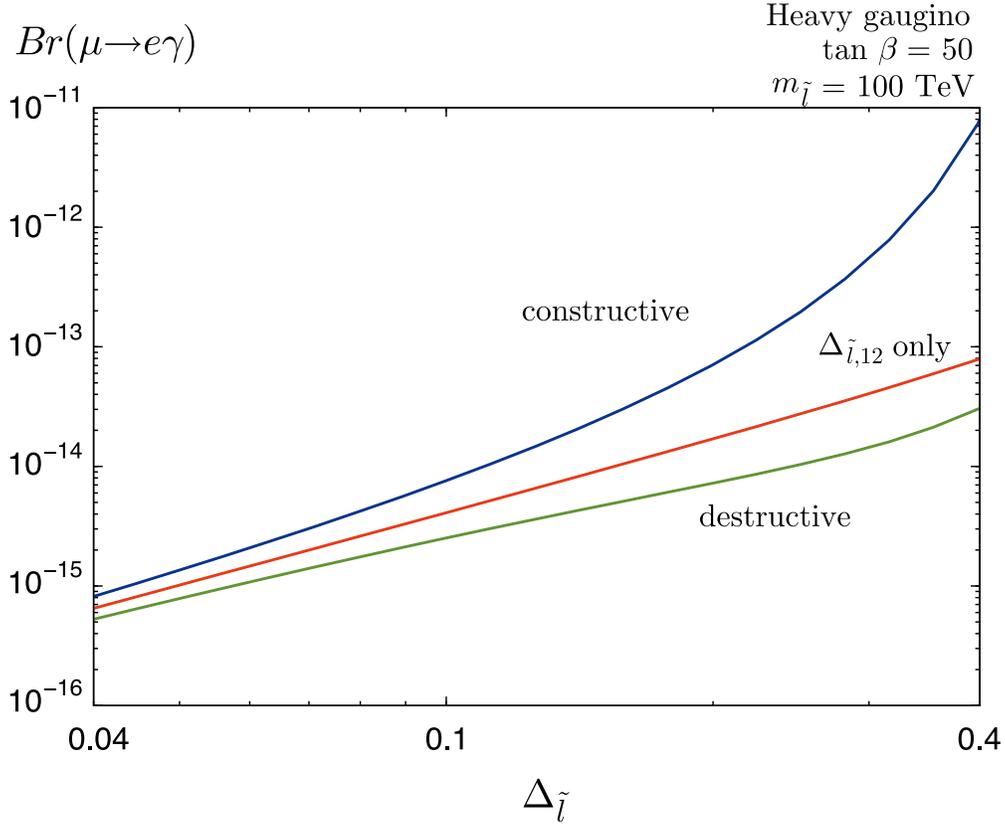}}
  \caption{$Br(\mu\rightarrow e\gamma$) as a function of the slepton
    mixing parameter $\Delta_{\tilde{l}}$ (which is the absolute value
    of the off-diagonal elements of $\Delta_{\tilde{l}_L,ij}$ and
    $\Delta_{\tilde{l}_R,ij}$) for the heavy gaugino scenario.  We
    take $\tan\beta=50$ and $\mu=m_{\tilde{l}}=100\ {\rm TeV}$. We
    show the constructive case
    ($\Delta_{\tilde{e}_R,12}=\Delta_{\tilde{e}_R,13}=\Delta_{\tilde{e}_R,23}=
    \Delta_{\tilde{l}_L,12}=\Delta_{\tilde{l}_L,13}=\Delta_{\tilde{l}_L,23}$)
    and destructive case
    ($-\Delta_{\tilde{e}_R,12}=\Delta_{\tilde{e}_R,13}=\Delta_{\tilde{e}_R,23}=
    -\Delta_{\tilde{l}_L,12}=\Delta_{\tilde{l}_L,13}=\Delta_{\tilde{l}_L,23}$).
    For comparison, we also show the results only with
    $\Delta_{\tilde{e}_R,12}$ and $\Delta_{\tilde{l}_L,12}$.}
  \label{fig:brMEG_delta}
\end{figure}

So far, we have considered the case where $\mu\rightarrow e\gamma$ is
dominantly induced by the 12 elements of $\Delta_{\tilde{l}_L,ij}$ and
$\Delta_{\tilde{l}_R,ij}$.  Other components may, however, also affect
$Br(\mu\rightarrow e\gamma)$.  In particular, $Br(\mu\rightarrow
e\gamma)$ can be enhanced if the product
$\Delta_{\tilde{e}_R,13}\Delta_{\tilde{l}_L,32}$ or
$\Delta_{\tilde{l}_L,13}\Delta_{\tilde{e}_R,32}$ are non-vanishing
\cite{Barbieri:1995tw}.  This is because, in such a case, left-right
mixing occurs due to the Yukawa interaction of tau-lepton instead of
that of muon.  To see how large the branching ratio can be, we also
calculated $Br(\mu\rightarrow e\gamma)$ for the case where the
absolute values of all the off-diagonal elements of
$\Delta_{\tilde{l}_L,ij}$ and $\Delta_{\tilde{l}_R,ij}$ are equal.
(Here, we take $\Delta_{\tilde{l}_L,ii}=\Delta_{\tilde{l}_R,ii}=0$.)
In such a case, the magnitude of the amplitude proportional to the tau
Yukawa coupling constant and that to the muon Yukawa coupling constant
become comparable when the off-diagonal elements are about $0.1$.  The
relative phase between those two amplitudes depends on the phases in
the off-diagonal elements.  In Fig.\ \ref{fig:brMEG_delta}, for the
fixed value of the slepton mass of $m_{\tilde{l}}=100\ {\rm TeV}$, we
plot $Br(\mu\rightarrow e\gamma)$ as a function of
$\Delta_{\tilde{l}}$ (which is the absolute value of the off-diagonal
elements of $\Delta_{\tilde{e}_R,ij}$ and $\Delta_{\tilde{l}_L,ij}$)
for the cases where the two amplitudes are constructive and
destructive.

Our results should be compared with the experimental bounds on the
leptonic flavor and CP violations as well as with the prospects of
future experiments (see Table \ref{table:experiments}).  As one can
see, the heavy sfermion scenario is already constrained by the present
bounds on $Br(\mu\rightarrow e\gamma)$ if slepton masses are below
$O(10\ {\rm TeV})$.  In the future, experimental bound on
$Br(\mu\rightarrow e\gamma)$ may be significantly improved by the MEG
upgrade, with which $\mu\rightarrow e\gamma$ may be found if the
branching ratio is larger than $\sim 6\times 10^{-14}$.  With the
choice of parameters used in Fig.\ \ref{fig:brMEG}, the MEG upgrade
will cover the slepton mass up to $\sim 55\ {\rm TeV}$, $43\ {\rm
  TeV}$, $25\ {\rm TeV}$ ($36\ {\rm TeV}$, $28\ {\rm TeV}$, $16\ {\rm
  TeV}$) for $\tan\beta=50$, $30$, and $10$, respectively, for the
heavy gaugino case (the AMSB case).

We also study $\mu\rightarrow 3e$ and $\mu$-$e$ conversion processes.
Here, we are paying particular attention to the case where $\tan\beta$
is large, with which the LFV rates are enhanced.  Then, dipole-type
operators give dominant contributions to the LFV processes because
their coefficients are proportional to $\tan\beta$.  In such a case,
we can use the following approximated formula to evaluate
$Br(\mu\rightarrow 3e)$ \cite{Ellis:2002fe}:
\begin{align}
  \frac{Br(\mu\rightarrow 3e)}{Br(\mu\rightarrow e\gamma)}
  \simeq 
  \frac{\alpha}{3\pi} 
  \left( 
    \log \frac{m_\mu^2}{m_e^2} - \frac{11}{4}
  \right)
  \simeq 6.6 \times 10^{-3},
\end{align}
where $\alpha$ is the fine structure constant.  In addition, in the
case of the dipole dominance, the $\mu$-$e$ conversion rate, which is
defined as
\begin{align}
  R_{\mu e} \equiv
  \frac{\Gamma(\mu N\rightarrow e N)}
  {\Gamma(\mu N\rightarrow {\rm capture})},
\end{align}
is also approximately proportional to $Br(\mu\rightarrow e\gamma)$ as
\begin{align}
  \frac{R_{\mu e}}{Br(\mu\rightarrow e\gamma)}
  \simeq 
  \frac{\pi D_N^2}
  {m_\mu^5 \tau_\mu \Gamma(\mu N\rightarrow {\rm capture})},
\end{align}
where $\tau_\mu$ is the lifetime of muon, and $D_N$ is the overlap
integral for the conversion process with nucleus $N$.  Using the
values of $D_N$ (with method 2) and capture rates given in
\cite{Kitano:2002mt}, the ratio $R_{\mu e}/Br(\mu\rightarrow e\gamma)$
is given by $2.5\times 10^{-3}$ and $3.0\times 10^{-3}$, for $N$ being
${}_{13}^{27}$Al and ${}_{\ 79}^{197}$Au, respectively.  We can see
that $\mu\rightarrow 3e$ and $\mu$-$e$ conversion processes may be
also within the reaches of future experiments.  (See Table
\ref{table:experiments}.)

\begin{table}[t]
  \begin{center}
    \begin{tabular}{lll}
      \hline \hline
      $\mu\rightarrow e\gamma$ 
      & MEG (current) \cite{Adam:2013mnn}$^\heartsuit$
      & $Br(\mu\rightarrow e\gamma) < 5.7\times 10^{-13}$
      \\
      & MEG Upgrade \cite{Baldini:2013ke}
      & $Br(\mu\rightarrow e\gamma) \lesssim 6\times 10^{-14}$ 
      \\ \hline
      $\mu\rightarrow 3e$
      & SINDRUM I \cite{Bellgardt:1987du}$^\heartsuit$
      & $Br(\mu\rightarrow 3e) < 1\times 10^{-12}$
      \\
      & Mu3e Phase I \cite{Blondel:2013ia}
      & $Br(\mu\rightarrow 3e) \lesssim 10^{-15}$
      \\
      & Mu3e Phase II \cite{Blondel:2013ia}
      & $Br(\mu\rightarrow 3e) \lesssim 10^{-16}$
      \\ \hline
      $\mu$-$e$ conversion
      & SINDRUM II \cite{Bertl:2006up}$^\heartsuit$
      & $R_{\mu e} < 7\times 10^{-13}$ (with Au)
      \\
      & DeeMe \cite{Aoki:2012zza}
      & $R_{\mu e} \lesssim 10^{-14}$ (with SiC)
      \\
      & Mu2e \cite{Abrams:2012er}
      & $R_{\mu e} \lesssim 2.4\times 10^{-17}$ (with Al)
      \\
      & COMET \cite{Kuno:2012pt}
      & $R_{\mu e} \lesssim 10^{-17}$ (with Al)
      \\
      & PRISM/PRIME \cite{Kuno:2012pt}
      & $R_{\mu e} \lesssim 2\times 10^{-19}$
      \\ \hline
      Electron EDM
      & YbF molecule \cite{Hudson:2011zz}$^\heartsuit$
      & $|d_e| < 10.5 \times 10^{-28} e\ {\rm cm}$
      \\
      & ThO molecule \cite{Vutha:2009ux}
      & $|d_e| \lesssim 3.7 \times 10^{-29} /\sqrt{D} e\ {\rm cm}$
      \\
      & Fr \cite{Sakemi:2011zz}
      & $|d_e| \lesssim 1 \times 10^{-29} e\ {\rm cm}$
      \\
      & YbF molecule \cite{Kara:2012ay}
      & $|d_e| \lesssim 1 \times 10^{-30} e\ {\rm cm}$
      \\
      & WN ion \cite{Kawall:2011zz}
      & $|d_e| \lesssim 1 \times 10^{-30} e\ {\rm cm}$
      \\
      \hline \hline
    \end{tabular}
    \caption{Current bounds on the leptonic flavor and CP violations
      (which are with ``$\heartsuit$''), as well as the future
      prospects.  ($D$ is the number of the days of operation.)}
    \label{table:experiments}
  \end{center}
\end{table}%

Before discussing the electron EDM, we comment on LFV decay processes
of $\tau$-lepton.  If $\Delta_{\tilde{l},23}$ and
$\Delta_{\tilde{l},12}$ are of the same order,
$Br(\tau\rightarrow\mu\gamma)$ becomes comparable to
$Br(\mu\rightarrow e\gamma)$.  Given the fact that even the BELLE II
experiment will reach $Br(\tau\rightarrow\mu\gamma)\sim 2.4\times
10^{-9}$ \cite{O'Leary:2010af}, it is harder to find the LFV processes
of $\tau$-lepton in models with heavy sfermions.

Next, let us consider the SUSY contribution to the electron EDM
$d_e^{\rm (SUSY)}$.  There are many sources of CP violation in the
present model; the phase in the $\mu$ parameter, those in the
off-diagonal elements of the sfermion mass matrices, and so on.  With
those, the SUSY contribution to the electron EDM may become sizable as
we see below. 

The SUSY contribution to the electron EDM can be generated even
without flavor violation.  However, with non-vanishing values of
$\Delta_{\tilde{e}_R,ij}$ and $\Delta_{\tilde{l}_L,ij}$, $d_e^{\rm
  (SUSY)}$ may be enhanced \cite{Barbieri:1995tw, Hisano:2008hn}.  
This is because the left-right mixing of
the smuon and/or stau, which are larger than that of selectron, may
contribute to $d_e^{\rm (SUSY)}$.  In the following, we show results
for the case with and without sizable flavor violations.  We calculate
one-loop SUSY diagrams contributing to the electron EDM, with slepton
and gaugino inside the loop.\footnote
{If the $\mu$ parameter is as small as gaugino masses, two-loop
  diagram may dominate the SUSY contribution to $d_e$
  \cite{ArkaniHamed:2004yi}.  In the present case, the $\mu$ parameter
  is as large as sfermion masses.}

\begin{figure}[t]
  \centerline{\epsfxsize=0.8\textwidth\epsfbox{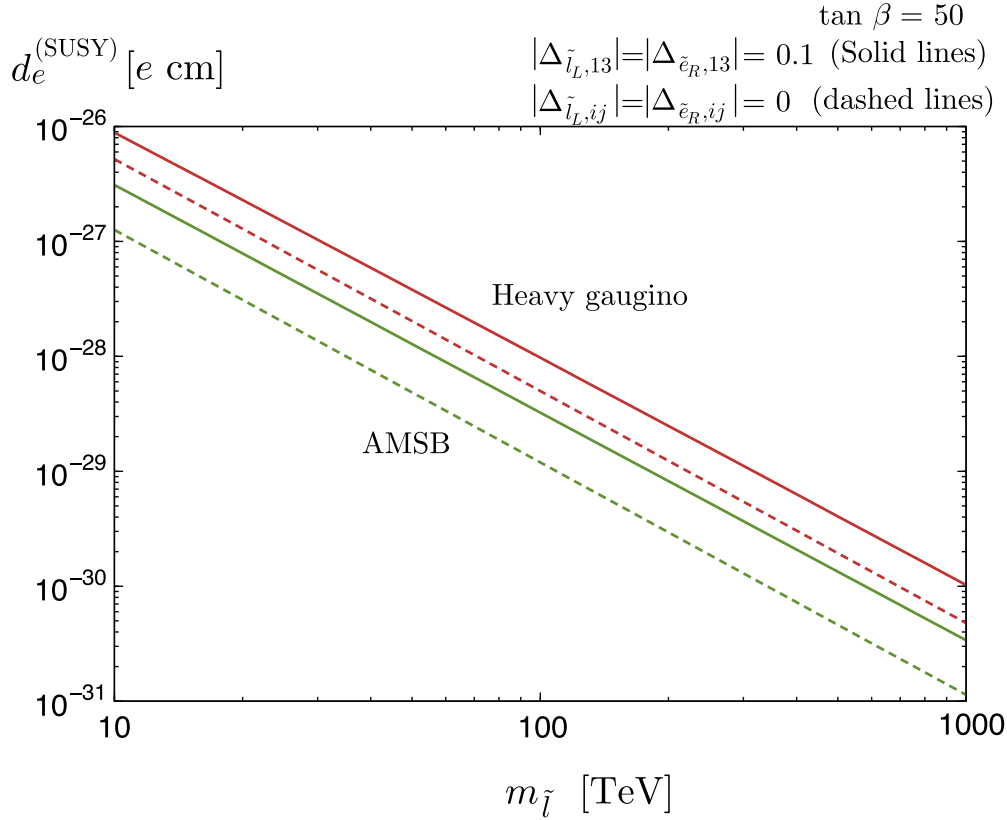}}
  \caption{The SUSY contribution to the electron EDM as a function of
    the slepton mass $m_{\tilde{l}}$ for $\tan\beta=50$ and
    $|\mu|=m_{\tilde{l}}$.  Upper (red) and lower (green) lines are
    for the heavy gaugino case with $M_3=m_{\tilde{l}}$ and the AMSB
    case with $m_{3/2}=5m_{\tilde{l}}$, respectively.  For the dashed
    lines, all the elements in $\Delta_{\tilde{e}_R,ij}$ and
    $\Delta_{\tilde{e}_l,ij}$ are taken to be zero, and the phase of
    $\mu$ is chosen to maximize $d_e^{\rm (SUSY)}$.  For the solid
    lines, $\Delta_{\tilde{e}_R,ij}=\Delta_{\tilde{l}_L,ij}=0$ except
    for $|\Delta_{\tilde{e}_R,13}|=|\Delta_{\tilde{e}_R,31}|=
    |\Delta_{\tilde{l}_L,13}|=|\Delta_{\tilde{l}_L,31}|=0.1$ and ${\rm
      Arg}(\mu)=0$; in this case, the phases of the off-diagonal
    elements are chosen to maximize $d_e^{\rm (SUSY)}$.}
   \label{fig:de}
\end{figure}

In Fig.\ \ref{fig:de}, we plot $d_e^{\rm (SUSY)}$ as a function of the
slepton mass $m_{\tilde{l}}$.  We consider the cases of heavy gauginos
and AMSB.  We adopt two cases of off-diagonal elements of the sfermion
mass matrices.  The first one is the case without flavor violation; we
take $\Delta_{\tilde{e}_R,ij}=\Delta_{\tilde{l}_L,ij}=0$.  In this
case, using the fact that the phase of $\mu$ can be arbitrary, we
choose ${\rm Arg}(\mu)$ which maximizes $d_e^{\rm (SUSY)}$.\footnote
{We use the convention such that the gaugino masses as well as the
  vacuum expectation values of the Higgs bosons are real.}
In addition, to see the effects of muon and tau Yukawa coupling
constants, we consider the second case in which $13$ and $31$
components of $|\Delta_{\tilde{e}_R,ij}|$ and
$|\Delta_{\tilde{l}_L,ij}|$ are $0.1$.  (Other components of
$\Delta_{\tilde{e}_R,ij}$ and $\Delta_{\tilde{l}_L,ij}$ are taken to
be zero.)  In this case, we assumed that the phase in $\mu$ is
negligible, and the phases of the off-diagonal elements are chosen
such that $d_e^{\rm (SUSY)}$ is maximized.  Notice that, in the second
case, $d_e^{\rm (SUSY)}$ is proportional to
$|\Delta_{\tilde{e}_R,13}\Delta_{\tilde{l}_L,31}|$ (with the phases of
off-diagonal elements being fixed).  As in the case of the LFV, the
electron EDM in the heavy gaugino case is larger than that in the AMSB
case.  In addition, $d_e$ is more enhanced as $\tan\beta$ becomes
larger; $d_e$ is approximately proportional to $\tan\beta$.

So far, we have concentrated on the leptonic flavor and CP violations.
However, it is well known that the SUSY models may also affect flavor
and CP violations of baryons.  It is often the case that
$K^0$-$\bar{K}^0$ mixing parameters, in particular the $\epsilon_K$
parameter, give very stringent constraints on the scale of the
superparticle masses \cite{Gabbiani:1996hi, Kersten:2012ed, Mescia:2012fg, Kadota:2011cr}.  To see
the importance of the constraints from the SUSY contribution to
$K^0$-$\bar{K}^0$ mixing parameters, we parameterize the mass matrices
as\footnote
{The right-handed up-type squarks are not important for the discussion
  of $K^0$-$\bar{K}^0$ mixing, so we simply take
  $\Delta_{\tilde{u}_R,ij}=0$.}
\begin{align}
  {\cal M}^2_{\tilde{d}_{R},ij} =&
  m_{\tilde{q}}^2 \left( \delta_{ij} + \Delta_{\tilde{d}_R,ij} \right),
  \\
  {\cal M}^2_{\tilde{d}_{L},ij} =&
  m_{\tilde{q}}^2 \left( \delta_{ij} + \Delta_{\tilde{d}_L,ij} \right),
  \\
  {\cal M}^2_{\tilde{u}_{R},ij} =&
  m_{\tilde{q}}^2 \delta_{ij},
\end{align}
and calculate the SUSY contribution.

\begin{figure}[t]
  \centerline{\epsfxsize=0.8\textwidth\epsfbox{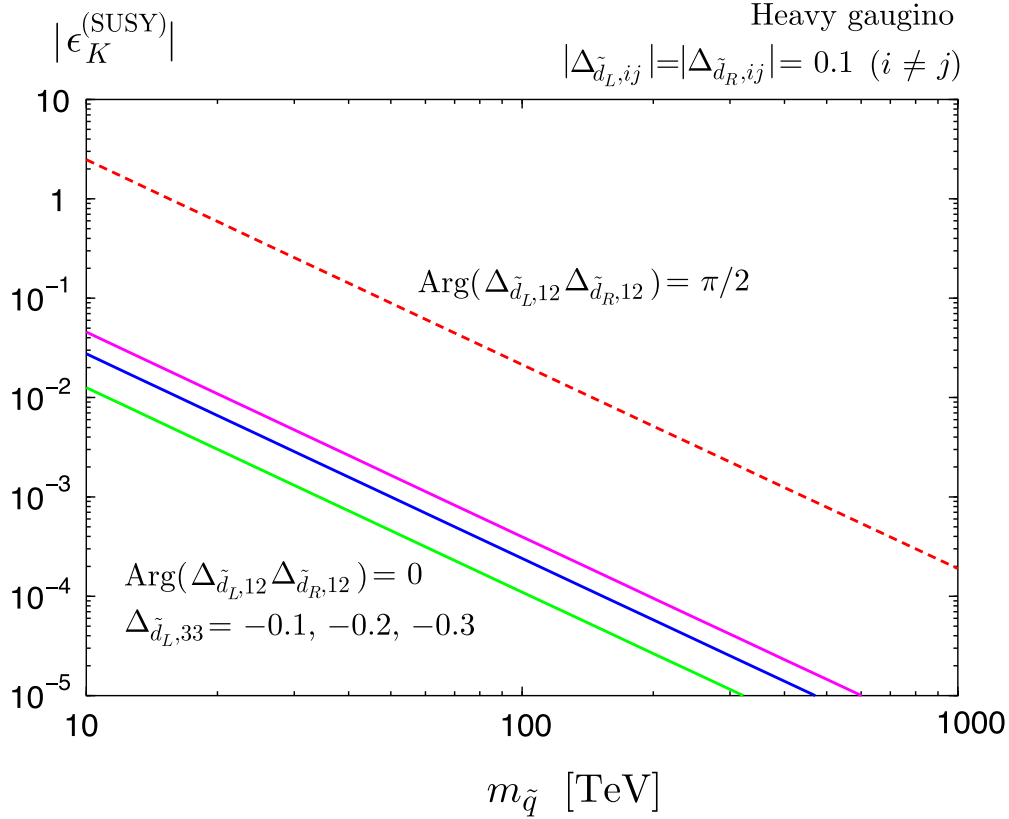}}
  \caption{The SUSY contribution to the $\epsilon_K$ parameter as a
    function of $m_{\tilde{q}}$.  Here, we consider heavy gaugino
    scenario taking $M_3=m_{\tilde{q}}$.  For the dashed line, we take
    $|\Delta_{\tilde{d}_R,ij}|=|\Delta_{\tilde{d}_L,ij}|=0.1$ ($i\neq
    j$) and ${\rm
      Arg}(\Delta_{\tilde{d}_R,12}\Delta_{\tilde{d}_L,12})=\pi/2$.
    For solid lines, we take $|\Delta_{\tilde{d}_R,ij}|=0.1$ ($i\neq
    j$) with the $SO(10)$ relation given in Eq.\ \eqref{so10}, except
    for $\Delta_{\tilde{d}_R,33}=0$ and
    $\Delta_{\tilde{d}_L,33}=-0.1$, $-0.2$, and $-0.3$ from below; the
    phases of the off-diagonal elements are chosen to maximize
    $\epsilon_K^{\rm (SUSY)}$.}
   \label{fig:epsk}
\end{figure}

In Fig.\ \ref{fig:epsk}, we plot the SUSY contribution
$\epsilon_K^{\rm (SUSY)}$ for the heavy gaugino case.  (We checked
that the result for the AMSB case does not change so much, and hence
the following arguments are also applicable to the AMSB case.)  In the
calculation of $\epsilon_K^{\rm (SUSY)}$, we use the formulae given in
\cite{Baek:2001kh}.  The dashed line shows the result for the case
where there is no cancellation; here, we take
$|\Delta_{\tilde{d}_R,ij}|=|\Delta_{\tilde{d}_L,ij}|=0.1$ ($i\neq j$)
and ${\rm Arg}(\Delta_{\tilde{d}_R,12}\Delta_{\tilde{d}_L,12})=\pi/2$.

Comparing the result with the bound on the possible extra contribution
to $\epsilon_K$, which we conservatively take $|\epsilon_K^{\rm
  (SUSY)}| =9.8\times 10^{-4}$,\footnote
{ We take into account the large uncertainty of the standard model
  contribution $\epsilon_K^{\rm (SM)} = (1.81\pm 0.28) \times 10^{-3}$
  \cite{Brod:2011ty}, while we use the experimental value
  $\epsilon_K^{\rm (exp)} = (2.228\pm 0.011) \times 10^{-3}$
  \cite{Nakamura:2010zzi}.}
the squarks are required to be heavier than $450\ {\rm TeV}$.  (Notice
that $\epsilon_K^{\rm (SUSY)}$ is approximately proportional to
$|\Delta_{\tilde{d}_R,12}\Delta_{\tilde{d}_L,12}|$ unless there is
cancellation.)  Even if the squark and slepton masses are of the same
order, the future electron EDM experiment may have a sensitivity to
such a parameter region if the bound on $d_e$ is improved by three
orders of magnitude (see Fig.\ \ref{fig:de}).

If the constraint from $\epsilon_K$ is relaxed, the future LFV
experiments also play important role to probe the heavy sfermion
scenario.  One observation is that $\epsilon_K^{\rm (SUSY)}$ is
suppressed if the following relation (which we call the $SO(10)$
relation because of the reason explained below) holds:
\begin{align}
  SO(10):~
  {\cal M}^2_{\tilde{d}_{R},ij}={\cal M}^{2*}_{\tilde{d}_{L},ij}.
  \label{so10}
\end{align}
This is from the fact that, if we limit ourselves to the sector
consisting of down-type (s)quarks and gluino, which gives the dominant
contribution to $\epsilon_K^{\rm (SUSY)}$, the Lagrangian becomes
invariant under the exchange of $d_L$ and $d_R^c$ (i.e., $C$
invariance).  Thus, if the relation \eqref{so10} is exact, the SUSY
contribution to $\epsilon_K$ comes from $SU(2)_L$ and $U(1)_Y$
interactions, resulting in a significant suppression of
$\epsilon_K^{\rm (SUSY)}$.

It is notable that the relation \eqref{so10} naturally arises from
$SO(10)$ unification of the gauge groups (at least at the GUT scale).
In $SO(10)$ model, $q_{L,I}$, $u_{R,I}^c$, $d_{R,I}^c$, $l_{L,I}$ and
$e_{R,I}^c$ (as well as right-handed neutrino) are embedded into a
single ${\bf 16}$-representation multiplet of $SO(10)$, which we
denote ${\bf 16}_I$.  In such a model, the MSSM Yukawa interactions
responsible for the quark and lepton masses arise from terms which are
quadratic in ${\bf 16}$-multiplet, and the Yukawa matrices in the
MSSM originate to symmetric $3\times 3$ matrices.\footnote
{Here, we assume that the effects of higher-dimensional operators are
  sub-dominant.}
The down-type Yukawa matrix in this basis can be diagonalized by a
single unitary matrix, which we denote $U_d$.  Then, the down-type
squarks in the flavor-eigenstate basis are related to
$\tilde{d}_{L,I}$ and $\tilde{d}_{R,I}^c$ (which are in the same ${\bf
  16}_I$ multiplet) as $\tilde{d}_{L,i}=U_{d,iJ}\tilde{d}_{L,J}$ and
$\tilde{d}_{R,i}^{c*}=U_{d,iJ}^* \tilde{d}_{R,J}^{c*}$, from which we
obtain the $SO(10)$ relation \eqref{so10}.

Even if the relation \eqref{so10} is satisfied at the GUT scale,
however, it may not hold at the lower energy scale.  In particular,
the RG effects change the relation.  The most important RG effect is
from the Yukawa coupling constants of third generation quarks (i.e.,
top and bottom quarks), with which the $33$ components of the mass
matrices of $\tilde{q}_L$ and $\tilde{d}_R^c$ are reduced.  Detailed
values of $\Delta_{\tilde{d}_L,33}$ and $\Delta_{\tilde{d}_R,33}$
depend on various parameters.  To see how large the difference may
become, we simply adopted the assumption of the universal scalar
masses at the GUT scale and estimated $\Delta_{\tilde{d}_L,33}$ and
$\Delta_{\tilde{d}_R,33}$ at the scale of $m_{\tilde{q}}$.  Then, we
found that
$(\Delta_{\tilde{d}_L,33},\Delta_{\tilde{d}_R,33})\simeq(0.7,1)$,
$(0.7,0.8)$, and $(0.5,0.5)$, for $\tan\beta=10$, $30$, and $50$,
respectively.\footnote
{One of the reasons why $\Delta_{\tilde{d}_L,33}$ and
  $\Delta_{\tilde{d}_R,33}$ are close to each other in particular for
  the large $\tan\beta$ case is that the SUSY breaking masses for up-
  and down-type Higgses are taken to be equal at the GUT scale.
}
We also calculate the SUSY contribution to $\epsilon_K$ for the case
where the relation \eqref{so10} holds except for $33$ components; we
show the results in Fig.\ \ref{fig:epsk} for the cases with
$\Delta_{\tilde{d}_R,33}\neq\Delta_{\tilde{d}_L,33}$.  Off-diagonal
elements are taken as
$\Delta_{\tilde{d}}=|\Delta_{\tilde{d}_R,ij}|=0.1$.  (In this case,
$\epsilon_K^{\rm (SUSY)}$ is approximately proportional to
$\Delta_{\tilde{d}}^3$ with
$\Delta_{\tilde{d}_R,33}-\Delta_{\tilde{d}_L,33}$ being fixed.)  As
one can see, the bound on $m_{\tilde{q}}$ is significantly relaxed in
this case in particular when $\Delta_{\tilde{d}_R,33}$ and
$\Delta_{\tilde{d}_L,33}$ are close.  In fact, in the case where
$\epsilon_K^{\rm (SUSY)}$ is suppressed, one should also consider the
constraint from the $K_L$-$K_S$ mass difference $\Delta m_K$.  We have
also calculated the SUSY contribution $\Delta m_K^{\rm (SUSY)}$, and
found that $m_{\tilde{q}}$ smaller than about $25\ {\rm TeV}$ is
excluded with the present choice of parameters by requiring that
$\Delta m_K^{\rm (SUSY)}$ should be smaller than the present
experimental value.

In summary, we have studied the leptonic flavor and CP violations in
supersymmetric models with heavy sfermions.  We have shown that the
SUSY contribution to the leptonic flavor and CP violations can be so
large that the future experiments may observe the signal even if
$m_{\tilde{l}}\sim O(10-100\ {\rm TeV})$.  The $\epsilon_K$ parameter
often gives a very stringent constraint on the mass scale of
superparticles if the off-diagonal elements of the sfermion mass
matrices are sizable.  However, it should be noted that the SUSY
contributions to the leptonic flavor and CP violations and those to
$K^0$-$\bar{K}^0$ mixing parameters depend on different parameters.
Thus, it is important to look for signals of new physics using
leptonic flavor and CP violation experiments.  In particular, we have
shown that, in some class of model like the $SO(10)$ unification
model, suppression of $\epsilon_K^{\rm (SUSY)}$ may occur because of
the automatic cancellation due to the approximate $C$ invariance.  In
addition, $\epsilon_K^{\rm (SUSY)}$ may be suppressed due to an
accidental cancellation.  In this letter, we have concentrated on the
leptonic sector (as well as the constraints from $K^0$-$\bar{K}^0$
mixing).  Other possible signals of the heavy sfermion scenario may be
hidden in the $B$ physics.  Such a possibility, as well as more
detailed studies of the leptonic flavor and CP violations, will be
given elsewhere \cite{MorNagYam}.

\vspace{1em}
\noindent {\it Acknowledgements}: The authors are grateful to Yasuhiro
Yamamoto for fruitful discussion and comments.  This work is supported
in part by Grant-in-Aid for Scientific research from the Ministry of
Education, Science, Sports, and Culture (MEXT), Japan, No.\ 22244021,
No.\ 23104008, and No.\ 60322997.


\begin{thebibliography}{99}

\bibitem{Aad:2012gk}
  G.~Aad {\it et al.}  [ATLAS Collaboration],
  Phys.\ Lett.\ B {\bf 716} (2012) 1.

\bibitem{Chatrchyan:2012gu}
  S.~Chatrchyan {\it et al.}  [CMS Collaboration],
  Phys.\ Lett.\ B {\bf 716} (2012) 30.
  
\bibitem{Aad:2012fqa}
  G.~Aad {\it et al.}  [ATLAS Collaboration],
  arXiv:1208.0949 [hep-ex].

\bibitem{Chatrchyan:2012jx}
  S.~Chatrchyan {\it et al.}  [CMS Collaboration],
  JHEP {\bf 1210} (2012) 018.

\bibitem{Kawasaki:2008qe}
  M.~Kawasaki, K.~Kohri, T.~Moroi and A.~Yotsuyanagi,
  Phys.\ Rev.\ D {\bf 78} (2008) 065011.

\bibitem{Giudice:2011cg}
  G.~F.~Giudice and A.~Strumia,
  Nucl.\ Phys.\ B {\bf 858} (2012) 63.

\bibitem{Ibe:2011aa}
  M.~Ibe and T.~T.~Yanagida,
  Phys.\ Lett.\ B {\bf 709} (2012) 374.

\bibitem{Moroi:2011ab}
  T.~Moroi and K.~Nakayama,
  Phys.\ Lett.\ B {\bf 710} (2012) 159.

\bibitem{Kane:2011kj}
  G.~Kane, P.~Kumar, R.~Lu and B.~Zheng,
  Phys.\ Rev.\ D {\bf 85} (2012) 075026.

\bibitem{ArkaniHamed:2012gw}
  N.~Arkani-Hamed, A.~Gupta, D.~E.~Kaplan, N.~Weiner and T.~Zorawski,
  arXiv:1212.6971 [hep-ph].

\bibitem{Hall:2013uga}
  L.~J.~Hall, J.~T.~Ruderman and T.~Volansky,
  arXiv:1302.2620 [hep-ph].

\bibitem{Ibe:2006de}
  M.~Ibe, T.~Moroi and T.~T.~Yanagida,
  Phys.\ Lett.\ B {\bf 644} (2007) 355.

\bibitem{Arganda:2012qp}
  E.~Arganda, J.~L.~Diaz-Cruz, A.~Szynkman and ,
  arXiv:1211.0163 [hep-ph].

\bibitem{Arganda:2013ve}
  E.~Arganda, J.~L.~Diaz-Cruz, A.~Szynkman and ,
  arXiv:1301.0708 [hep-ph].

\bibitem{Gabbiani:1996hi}
  F.~Gabbiani, E.~Gabrielli, A.~Masiero and L.~Silvestrini,
  Nucl.\ Phys.\ B {\bf 477} (1996) 321.

\bibitem{Giudice:1988yz}
  G.~F.~Giudice and A.~Masiero,
  Phys.\ Lett.\ B {\bf 206} (1988) 480.

\bibitem{Inoue:1991rk}
  K.~Inoue, M.~Kawasaki, M.~Yamaguchi and T.~Yanagida,
  Phys.\ Rev.\ D {\bf 45} (1992) 328.

\bibitem{Randall:1998uk}
  L.~Randall and R.~Sundrum,
  Nucl.\ Phys.\ B {\bf 557} (1999) 79.

\bibitem{Giudice:1998xp}
  G.~F.~Giudice, M.~A.~Luty, H.~Murayama and R.~Rattazzi,
  JHEP {\bf 9812} (1998) 027.

\bibitem{Moroi:1999zb}
  T.~Moroi and L.~Randall,
  Nucl.\ Phys.\ B {\bf 570} (2000) 455.

\bibitem{Banks:1993en}
  T.~Banks, D.~B.~Kaplan and A.~E.~Nelson,
  Phys.\ Rev.\ D {\bf 49} (1994) 779.

\bibitem{Hisano:1995cp}
  J.~Hisano, T.~Moroi, K.~Tobe and M.~Yamaguchi,
  Phys.\ Rev.\ D {\bf 53} (1996) 2442.

\bibitem{Barbieri:1995tw}
  R.~Barbieri, L.~J.~Hall and A.~Strumia,
  Nucl.\ Phys.\ B {\bf 445} (1995) 219.

\bibitem{Ellis:2002fe}
  J.~R.~Ellis, J.~Hisano, M.~Raidal and Y.~Shimizu,
  Phys.\ Rev.\ D {\bf 66} (2002) 115013

\bibitem{Kitano:2002mt}
  R.~Kitano, M.~Koike and Y.~Okada,
  Phys.\ Rev.\ D {\bf 66} (2002) 096002
   [Erratum-ibid.\ D {\bf 76} (2007) 059902].



\bibitem{Adam:2013mnn}
  J.~Adam {\it et al.}  [MEG Collaboration],
  arXiv:1303.0754 [hep-ex].

\bibitem{Baldini:2013ke}
  A.~M.~Baldini {\it et al.},
  arXiv:1301.7225 [physics.ins-det].

\bibitem{Bellgardt:1987du}
  U.~Bellgardt {\it et al.}  [SINDRUM Collaboration],
  Nucl.\ Phys.\ B {\bf 299} (1988) 1.

\bibitem{Blondel:2013ia}
  A.~Blondel {\it et al.},
  arXiv:1301.6113 [physics.ins-det].

\bibitem{Bertl:2006up}
  W.~H.~Bertl {\it et al.}  [SINDRUM II Collaboration],
  Eur.\ Phys.\ J.\ C {\bf 47} (2006) 337.

\bibitem{Aoki:2012zza}
  M.~Aoki [DeeMe Collaboration],
  AIP Conf.\ Proc.\  {\bf 1441} (2012) 599.

\bibitem{Abrams:2012er}
  R.~J.~Abrams {\it et al.}  [Mu2e Collaboration],
  arXiv:1211.7019 [physics.ins-det].

\bibitem{Kuno:2012pt}
  Y.~Kuno,
  Nucl.\ Phys.\ Proc.\ Suppl.\  {\bf 225} (2012) 228.

\bibitem{Hudson:2011zz}
  J.~J.~Hudson {\it et al.},
  Nature {\bf 473} (2011) 493.

\bibitem{Vutha:2009ux}
  A.~C.~Vutha {\it et al.},
  J.\ Phys.\ B {\bf 43} (2010) 074007.

\bibitem{Sakemi:2011zz}
  Y.~Sakemi {\it et al.},
  J.\ Phys.\ Conf.\ Ser.\  {\bf 302} (2011) 012051.

\bibitem{Kara:2012ay}
  D.~M.~Kara {\it et al.},
  New J.\ Phys.\  {\bf 14} (2012) 103051.

\bibitem{Kawall:2011zz}
  D.~Kawall,
  J.\ Phys.\ Conf.\ Ser.\  {\bf 295} (2011) 012031.

\bibitem{O'Leary:2010af}
  B.~O'Leary {\it et al.}  [SuperB Collaboration],
  arXiv:1008.1541 [hep-ex].

\bibitem{Hisano:2008hn}
  J.~Hisano, M.~Nagai and P.~Paradisi,
  Phys.\ Rev.\ D {\bf 80} (2009) 095014.

\bibitem{ArkaniHamed:2004yi}
  N.~Arkani-Hamed, S.~Dimopoulos, G.~F.~Giudice and A.~Romanino,
  Nucl.\ Phys.\ B {\bf 709} (2005) 3.

\bibitem{Kersten:2012ed}
  J.~Kersten, L.~Velasco-Sevilla and ,
  arXiv:1207.3016 [hep-ph].

\bibitem{Mescia:2012fg}
  F.~Mescia, J.~Virto and ,
  Phys.\ Rev.\ D {\bf 86} (2012) 095004 .

\bibitem{Kadota:2011cr}
  K.~Kadota, G.~Kane, J.~Kersten and L.~Velasco-Sevilla,
  Eur.\ Phys.\ J.\ C {\bf 72} (2012) 2004.

\bibitem{Baek:2001kh}
  S.~Baek, T.~Goto, Y.~Okada and K.~-i.~Okumura,
  Phys.\ Rev.\ D {\bf 64} (2001) 095001.

\bibitem{Brod:2011ty}
  J.~Brod and M.~Gorbahn,
  Phys.\ Rev.\ Lett.\  {\bf 108} (2012) 121801.

\bibitem{Nakamura:2010zzi}
  K.~Nakamura {\it et al.}  [Particle Data Group Collaboration],
  J.\ Phys.\ G {\bf 37} (2010) 075021.

\bibitem{MorNagYam}
  T.~Moroi, M.~Nagai and Y.~Yamamoto,
  in preparation.


\end{thebibliography}
\end{document}